# Gamma-ray bursts at extremely small fluence

Vladimir Lipunov[a,*], Sergey Svertilov[a], Vladislav Topolev[a],

[a] Lomonosov Moscow State University

ABSTRACT

In this review we show that the space experiment with gamma-ray detector with sensitivity 2 orders of magnitude higher than existing ones will make it possible to discover up to a thousand neutron star mergers, even at those moments when gravitational wave (GW) antennas are not working. At the same time, synchronous detection of neutron stars mergers by gamma-ray and GW detectors will make it possible not only to study in detail the physical processes occurring at the time of the catastrophe, but also to determine the full gamma ray beam pattern, including the average jet divergence angle and the real energy of the explosion. A gamma detector that has the required sensitivity at a relatively low flight weight is proposed. The latter, in turn, will make it possible to clarify our ideas about the genesis of double relativistic stars in the Universe.

## 1. Introduction

The cosmological nature of gamma-ray bursts (GRBs) remained a mystery for more than a quarter of a century after their discovery, when their optical localization was first possible (van Paradijs et al., 1997, Costa et al., 1997). After this, it became clear that we are dealing with the most powerful electromagnetic phenomena in the Universe. By the way, cosmological models gave a natural explanation of "non-uniform" distribution of GRB sources, on which the dependence of the number of gamma-ray bursts ($N$) on their fluence ($S$), so-called $\log N$ - $\log S$ curve, was indicated (Mazets et al. 1981). As a result of the research, it was revealed that among GRBs two populations can be distinguished according to their duration (Kouveliotou et al., 1993), i.e. short (sGRBs) and long (lGRBs). The dominant explanation for the mechanism of occurrence of short bursts was the hypothesis of the merging of relativistic compact objects, i.e. two neutron stars or a neutron star and a black hole (Blinnikov et al. 1984, Lipunov et al. 1995a, Paczynski et al. 1998). Long bursts are usually associated with the collapse of the rapidly rotating core of a massive star (MacFadyen & Woosley, 2001; Lipunov & Gorbovskoy, 2007) but could also be produced by compact object mergers (e.g. GRB 060614 (Gehrels et al., 2006), GRB 211211A (Troja et al., 2022), GRB 230307A (Yang et al., 2024)).

According to observational data, less than a third of the bursts are sGRBs, lasting less than two seconds. For a long time, the study of sGRBs progressed poorly; only in 2005 the first afterglow associated with sGRB was detected. Further localization of several afterglows showed that, unlike lGRBs, which were highly correlated with star formation regions, some short bursts were associated with weak star formation regions or they were observed in regions without star formation, such as large elliptical galaxies (Gehrels et al., 2005).

By this, the mentioned above hypothesis of relativistic compact object merging has become dominant for sGRB explanation. Such mergers should be associated with star formation regions, but theoretical calculations suggested the possibility of similar events in fairly old systems.

There were several reasons to expect that neutron star mergers should be accompanied by electromagnetic radiation. After a merger (Clark et al., 1979), some of the neutron star's material can be ejected, and this can lead to radioactive decay of the synthesized heavy elements, i.e. so-called kilonova (Li & Paczynski, 1998; Tanvir et al., 2013). The origin of sGRBs during the neutron star mergers was confirmed when the short gamma-ray burst GRB170817A was detected less than two seconds after the detection of the gravitational wave GW170817, which was a signal from the two neutron stars merging (Abbott et al. 2017ab, Lipunov et al. 2017ab).

The discovery of gravitational waves has provided a new channel for exploring the Universe (Abbott et al. 2016a, b). Most of the events detected in the O1 (2015), O2 (2017), O3 (2019/20) periods of LIGO/Virgo operation were associated with black hole mergers (BHBH), which coincided with the Machine Scenario (Lipunov et al., 1997) prediction. But more interesting from the point of the GRB research, was the observation of neutron star merging on August 17, 2017 (Abbott et al. 2017a,b, Lipunov et al. 2017b, 2022), i.e. detection of gravitational wave GW170817, associated with the GRB170817A gamma-ray burst, which proved that at least some short gamma-ray bursts are indeed caused by neutron star mergers (Abbott et al. 2017a).

GW 170817 event occurred extremely close, at a distance of only about ~ 40 Mpc. However, it was this distance that appeared in the Machine Scenario calculations as the most probable among the first detections of neutron star mergers (Lipunov, Nazin, Panchenko et.al., 1995b; Curran & Lorimer, 1995; van den Heuvel & Lorimer, 1996). Consequently, GRB170817A has a very low isotropic luminosity of $\sim 10^{47}$ erg/s, which is four orders of magnitude lower than that observed for most other sGRBs.

Such a low isotropic luminosity has led to the hypothesis that there is a broader component of GRB emission than a typical jet, which, due to its low brightness, can only be observed from extremely short distances. In fact, this discovery showed that all gamma-ray observatories existing before 2017 would never "see" a neutron star merger at a distance of more than ~70 Mpc. This circumstance means that there is a urgent need to increase the sensitivity of space gamma-burst monitors by 2-3 orders of magnitude. This process has already begun in a number of projects (for example, such as GECAM (Ma 2019, Chen et al. 2020)). A very interesting question arises: what will these ultrasensitive gamma-ray observatories see and, most importantly, what new information about gamma-ray bursts and the Universe itself can numerous observations of GRBs provide?

This is the main goal of our study. It is obvious that increasing the sensitivity of gamma detectors by 2-3 orders of magnitude will make it possible to detect neutron star mergers at distances 10 - 30 times further, up to 500 - 1500 Mpc. In this way, we can provide synchronous gamma-ray and gravitational wave (GW) observations of the prompt emission of many thousands of colliding neutron stars per year. It is clear that such detectors should have wide fields of view (FOVs), which naturally will not significantly increase the accuracy of determining the coordinates of events and the error boxes will still remain at the GW detector level. Obviously, in this case we will be dealing with a large number of GRBs and, in a manner, we will find ourselves in a situation with GRBs before 1997, although with the important difference that we will have information about the distances to GRB sources, and in some cases with precise optical coordinates. In any case, at the first stages it will be possible to simply restore the multichannel nature of observations of GW sources, and it will certainly be possible to conduct statistical studies as during the era of the KONUS and Compton Observatory experiments (Aptekar et al., 1997, Schoenfelder et al. 1993).

Then we will focus on the simplest, and therefore reliable, predictions based on statistical modeling with the use of velocity data and estimates of the neutron star merging frequency, i.e. of the progenitors of sGRBs, that can predict the possible behavior of the log$N$ - log$S$ curve in a region beyond the sensitivity limits available at the moment.

Studying the behavior of the log$N$ - log$S$ curve for different parameters of the sGRB progenitors, inside and outside the sensitivity limit of the detectors, will allow, with the improvement of observational instruments, to quickly make a prediction regarding the possibility of the existence of a broader component in sGRB population. Comparison of typical observed bursts with a known redshift with GRB170817A allows us to obtain an estimation of the sGRB fraction observed outside the jet in current observations. Also, within the framework of the model, some limitations may be obtained on the possible variants of the connections, such as jet openings versus rate of neutron star mergers. All log$N$ - log$S$ curves for sGRBs are based on ten years of sGRB observations carried out by the Fermi GBM.

## 2. GRB observations

The first detected GRB is considered to be a burst of gamma-rays in the energy range of photons 0.1 - 1 MeV detected by the Vela military satellites in July 1967. The high frequency of events was striking, i.e. bursts were detected almost every day, which was first established in the excellent Soviet experiment KONUS, carried out under the PI E.P. Mazets on the Venera-11, -12 and Prognoz spacecraft in the 70s. For a long time, the origin of these bursts remained a mystery; even intra-galactic scenarios were suggested. Analysis of their energy spectra showed the non-thermal nature of the emission, as well as the strong magnetic field of the GRB progenitors, which suggested their connection with such compact objects as neutron stars. Improvements in the detector sensitivity in the 90s allowed observations with instruments such as BATSE CGRO, which show the isotropy of GRB sources on the sky with sufficiently high reliability (first indications on the isotropic GRB sources distribution on the sky were obtained just in experiments on Venera spacecraft), which forced us to reject the intra-galactic scenario of the GRB origin.

Significant for the further development of the GRB study was the discovery in 1997 of the GRB 970228, for which the afterglow in X-rays and optics was detected (van Paradijs et al., 1997, Costa et al., 1997), which was predicted in the work of Paczyński et al., 1993, and the host galaxy was discovered on z = 0.695 (Bloom, Djorgovski, Kulkarni, 2001). This observation definitively proved the extragalactic origin of the GRBs.

The extragalactic nature of GRBs established a new problem, i.e. the enormous value of isotropic luminosity, more than $10^{52}$ erg, which was later explained by the idea that the object releases energy only in a small solid angle, that is, the burst integral energy is several orders of magnitude less than the isotropic one (Sari, Piran, Halpern, 1999; Rhoads, 1999;). The narrow emission area is called a GRB jet. From this hypothesis a simple conclusion can be done, i.e. to detect a burst, the observer must be in a relatively small cone of emission, while most GRBs "shoot" past the observer. Later, it was also found that in the GRB distribution on duration, two typical groups with a boundary of 2 seconds can be distinguished, i.e. short and long GRBs (Kouveliotou et al., 1993), which probably should have had different origins.

### 2.1. GRB classification

The GRB light curves have a complicated structure and, for the most part, cannot be classified often they can be characterized as random. The duration of the observed emission can vary from milliseconds to thousands of seconds, the curve can consist of many sharp maxima or a pair of smoothed pulses in which there may or may not be symmetry. Some bursts are accompanied by a weaker precursor event, which is called a precursor. It occurs a few seconds or minutes before the start of the main burst; GRBs also exhibit an afterglow caused by a collision between burst emission and the interstellar matter, which, unlike the GRBs themselves, is already well described by some models (van Eerten, MacFadyen, 2012).

Although there are some simplified models that allow us to identify groups with similar behavior (Lipunov et al., 2017a), they consider only a very small amount of the total number of bursts and cannot explain the details of what is happening. However, the distributions of observed durations for a large number of GRBs show two distinct modes in the distribution (see Fig. 1, left panel). It is immediately worth noting here that due to the chaotic nature of the GRB curve, a convenient and applicable definition of duration is the time at which the gamma-ray burst has emitted 90% of its total energy, starting from the moment it exceeds 5% of the total energy, ending with reaching 95%. This definition will be used throughout this paper and in some places, it will be signed as $T_{90}$.

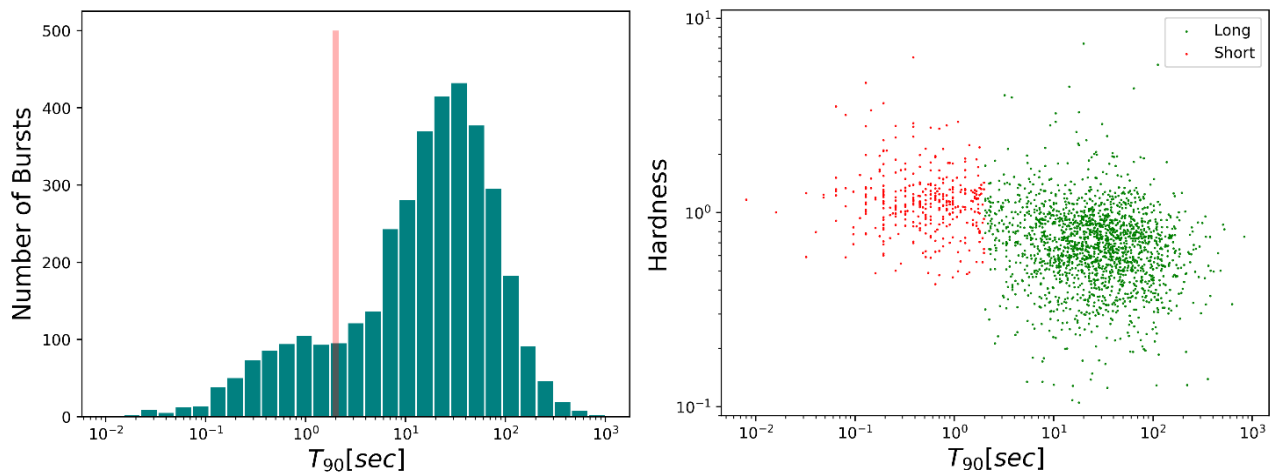

Fig. 1. Distribution on burst duration obtained by Fermi (von Kienlin et al., 2020), and scatter plots of spectral hardness vs. duration.

Based on the bimodal structure of the time distribution, it was supposed that there are two separate populations with different GRB emission process physics. Due to that such populations were identified primarily by time boundaries, they were named "short" with an average duration of about 0.3 s and "long" with an average duration of about 30 s. Both distributions are very broad and have a significant area of overlap, which makes it not always clear to classify a particular event based on duration alone. It is also necessary to note that, on average, short bursts have a harder spectrum than long bursts (Dezalay et al., 1991), see Fig. 1, right panel.

## 2.2. Long GRBs

Most observed GRBs last longer than two seconds and are classified as "long". Since long bursts are the majority, and they often have brighter afterglows, their properties have been studied in much more detail than those of short bursts. Long GRBs (lGRBs) are observed in galaxies with fast star formation (Bloom, Kulkarni, Djorgovski, 2002) and are often associated with Ic supernova explosions (Galama et al. 1998, Hjorth et al. 2003, Woosley and Bloom, 2006), which clearly associates long-lasting GRBs with the death of massive stars.

To explain lGRBs, several hypotheses have been proposed in which they are associated with the collapse of the rapidly rotating core of a massive star. Rapid rotation slows down the collapse and increases the time available for the creation of electromagnetic emission, which allows the prompt phase to last long enough.

In a more detailed description of the lGRB occurrence, several approaches are possible. One of the dominant scenarios is proposed by MacFadyen, Woosley 2001, where first a black hole is formed, and then a part of the supernova shell, which has a supercritical rotational moment, forms a heavy super-dense disk. Due to the generation of magnetic fields by this disk, axial jets with a large Lorentz factor are generated, which is observed. In another scenario, a rapidly rotating magnetized object, a spinar, is formed, which slowly shrinks due to torque dissipation (Lipunov & Gorbovskoy 2007). In this case, a jet with an energy flux determined by the Umov-Poynting vector is formed along the rotation axis. The operating time of the central GRB engine is determined by the momentum dissipation rate. In general, both scenarios require a fairly large torque in the collapsing stellar core. And here a duality scenario comes to mind for explanation, in which rapid rotation occurs due to the tidal influence of the second component in a very close binary system (Tutukov & Cherepashchuk 2016). As you can see, the exact mechanism of lGRB formation remains controversial.

## 2.3. Short GRBs

A minority of GRB last less than two seconds and are classified as short GRBs. Their part is less than a third of the total burst number. For a long time, the study of sGRBs progressed poorly; only in 2005 was the first afterglow from a short burst detected (Gehrels et al., 2005). Further localization of several afterglows showed that, unlike lGRBs, which were highly correlated with star formation areas, some short ones were associated with regions of weak star formation or were observed in areas without it at all, such as in large elliptical galaxies (O'Connor et al. 2022; Fong et al., 2022).

In this regard, for sGRBs, the dominant hypothesis has become the merging of relativistic compact objects, i.e. two neutron stars or a neutron star and a black hole. Such mergers should be associated with star formation areas, but theoretical calculations suggested the possibility of mergers in fairly old systems. We will discuss the function that describes the evolution of the neutron star merging rate later, but we will immediately note that when star formation is "turned off," the rate of mergers should fall. The simplest and most frequently used approximation of this falling with time is the law $\sim t^k$, where the value $k \sim 1$.

There were several reasons to expect that neutron star mergers should be accompanied by electromagnetic emission. After a merger, some of the neutron star's material can be ejected (Clark et al., 1979), and this can lead to radioactive decay of the synthesized heavy elements - the so-called kilonova (Li & Paczynski, 1998; Tanvir et al., 2013). The origin of sGRBs during neutron star mergers was confirmed when the sGRB GRB170817A was detected less than two seconds after the detection of the gravitational wave GW170817, which was a signal from the merging of two neutron stars. We 'll talk about event 170817 in more detail in the next section.

## 3. Events GW170817 and GRB170817A

The joint detection of GW GW170817 (Abbott et al. 2017a) and the similar sGRB GRB170817, followed by the detection of a kilonova (Lipunov et al., 2017b; McCully et al., 2017; Evans et al., 2017; Troja et al., 2017; Tanvir et al., 2017; Pian et al., 2017; Coulter et al., 2017; Arcavi et al., 2017; Valenti et al., 2017; Drout et al., 2017) not only strengthened the belief that sGRBs are formed from binary neutron star (BNS) mergers, but also ushered in an era of using GW observations together with multi-wavelength electromagnetic astronomy. The combination of data from such observations makes it possible to test existing models and refine their parameters. Thus, based on the observations carried out, the cosmological model parameters were studied, the principles of general relativity were verified, etc. (Abbott et al., 2017b).

However, the detection of a GW event not only provided answers, but also raised questions, since it occurred extremely close, at a distance of only about 41 Mpc (Abbott et al., 2019) ($z \approx 0.01$), giving an isotropic brightness at the source of only about $10^{47}$ erg/s, which is three orders of magnitude lower than the typical value observed for other sGRBs.

The GW data for this event alone cannot exclude the presence of objects more compact than neutron stars, such as black holes or, for example, hypothetical quark stars. Thus, observations in other bands help us to definitively identify the object.

Studies of signal properties based on merger models made it possible to obtain several more properties of the source that generated GW170817. Thus, the angle between the axis of the total angular momentum of the system and the line of sight was measured, and it was ~56°, and the masses of the components, $m_1$ and $m_2$, respectively, are in the range $m_1 \in (1.36 - 2.26)M_\odot$ and $m_2 \in (0.86 - 1.36)M_\odot$. The total mass was $M \in 2.82^{+0.47}_{-0.09}M_\odot$ (Abbott et al., 2017a). These results are consistent with a model of the merger of a two neutron star system. The event observation at a fairly wide angular distance from the central axis (Abbott et al., 2019) led to the hypothesis that the GRB170817A jet had some wide-angle structure with a lower isotropic luminosity, while it is likely that this object also had a typical central jet component.

Afterglow observations and further analysis confirm these conclusions (Troja et al., 2017, 2018, 2020). Quite a few attempts have been made to explain and model the GRB jet structure, but, unfortunately, at the moment it is difficult to identify the most plausible model (Huang et al., 2002; Xie, MacFadyen, 2019; Granot et al., 2018; Ryan et al., 2020). And although more detailed modeling of the jet structure should allow us to eliminate systematic distortions, the question of the jet structure for GRB170817 cannot be solved definitely, this is discussed in more detail below. Firstly, let us consider in a little more detail the observation of event 170817 in the gamma ray range.

The first GRB 170817A announcement was an automatically generated GCN message from the Fermi-GBM instrument (Fermi-GBM 2017, GCN, 524666471) just 14 s after the GRB detection at $T_0 = 12:41:06$ UTC. GRB 170817A was also detected by the INTEGRAL spacecraft with the SPI instrument, after receiving a signal from gravitational wave detectors.

Standard follow-up analyzes determined the pulse duration to be $T_{90} = 2.0 \pm 0.5$ s in the energy range 50 - 300 keV (Goldstein et al., 2017). As you can see, $T_{90}$ value corresponds the border of the two GRB groups, i.e. sGRBs and lGRBs. A more detailed analysis classified GRB170817A as sGRB with a probability of 73%. The classification of GRB170817A as sGRB is confirmed additionally by taking into account spectral hardness.

The sGRB had a peak photon flux, measured on a 64 ms time scale, of $3.7 \pm 0.9$ photons/s$^{-1}$cm$^{-2}$, and a flux of $(2.8 \pm 0.2)$ $10^{-7}$ erg cm$^{-2}$ s$^{-1}$ in the range 10 - 1000 keV (Goldstein et al., 2017), which is comparable to other bursts.

The announcements of the Fermi-GBM and LIGO-Virgo (GCN, 21505, GCN, 21527) observations kicked-on a campaign of broadband observations in search of electromagnetic identifications. A large number of teams around the world were mobilized using ground-based and space-based telescopes that could observe the area of GRB source location. August 17 at 23:33 UTC (+10.87 h from GRB event) with the use of 1-meter Swope telescope at Las Campanas Observatory in Chile (Gehrels et al., 2016), a transient was detected at a distance of 10 arcs from the center of NGC 4993, an early-type galaxy in the ESO 508 group, which is at a distance of 40 Mpc. Subsequently, several more teams independently observed this transient, including MASTER at 05:38 UTC the next day (Lipunov et al., 2017b).

Using spectral information and taking into account the distance to the host galaxy NGC 4993 $42.9 \pm 3.2$ Mpc, it was possible to find the energy of GRB170817A (Abbott B. P., et al., 2017c). The isotropic energy release in the gamma range was $E_{iso} = 3 \cdot 10^{46}$ erg, and the peak brightness was $L_{iso} = 1.6 \cdot 10^{47}$ erg/s in the energy range from 1 keV up to 10 MeV. If these isotropic values are compared with even the faintest of other sGRBs (Zhang et al., 2018), the differences are still greater.

Several scenarios have been proposed to explain this low luminosity (see Fig. 2):

1) the observer looks at the jet from a point outside the opening angle of the standard jet model, we are observing the so-called orphan burst (see Fig. 2, left picture);

2) the jet has a more complex structure than a cylinder with constant brightness, this structure makes it possible to observe emission at large angles with low luminosity (see Fig. 2, right picture);

3) there is some unique feature observed and this burst is not a typical GRB;

4) a unique feature is in the innermost emission, but the object is still essentially a sGRB.

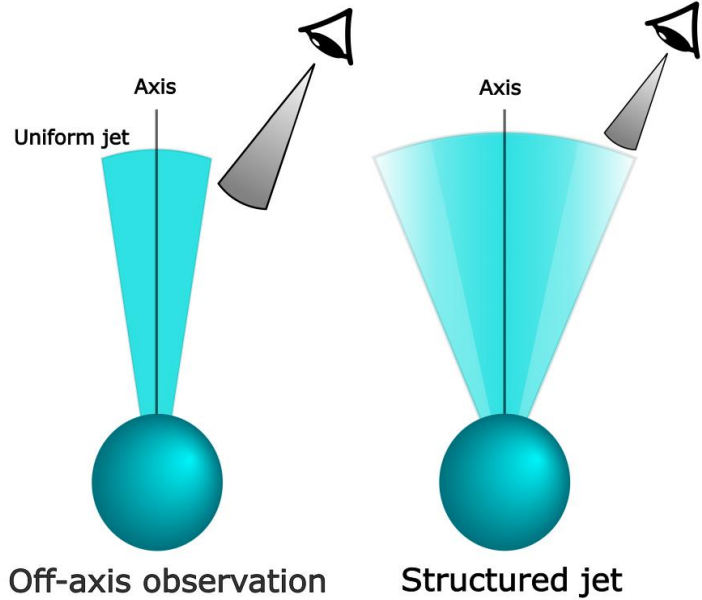

Fig. 2. Visualization of the first two scenarios for explaining low isotropic luminosity

**First variant.** A uniform cone-shaped jet (constant luminosity and Lorentz factor ($\Gamma$), within the jet opening angle) with a sharp edge, observed off-axis, where the luminosity should fall with some power law from the viewing angle. This leaves the question of a discontinuity in the observed values of isotropic luminosity, which would then have to be explained by a sharp slope in the dependence of luminosity on angle during off-axis observation. This model looks realistic, but as a first approximation such an off-axis observation should still lead to the appearance of a new group of observed sGRBs with a wider opening angle, but with a lower average brightness.

**Second variant.** The complex structure of the jet can naturally explain the wide observed fluence distribution (Ryan et al., 2023). But the lack of observations of intermediate variants makes us think that any model of a structured jet will be reduced to the need for the appearance of a new subgroup of close and weak bursts, that is, the jet structure should imply an abrupt transition.

**Third variant.** Taking into account the low distance to this burst source, it is possible that the observed emission is due to a mechanism distinct from sGRB, which is inherently typical to a faint bursts and therefore has not yet been detected. This seems unlikely since the main emission episode of GRB170817A shows all the properties of a typical sGRBs except the required energetics. This may probably be a general feature of faint short bursts, but then we again return to the observed bimodality in the distribution of isotropic luminosity, which cannot yet be studied in detail due to the lack of sensitivity of gamma-ray observatories.

**Fourth variant.** If GRB170817A is observed within a jet, and its dimness is explained by internal factors, we can conclude that sGRB jets have a distribution spanning at least six orders of magnitude, which seems impossible given the strong restrictions on the progenitor systems, i.e. NS + NS.

It is necessary to note that the X-ray emission from GW170817/GRB170817A event, which occurs with a delay relative to the GRB itself, may also indicate that this event observed off-axis (Mészáros et al., 1998; Granot et al., 2002). Thus, we can conclude that the most likely explanation for the low brightness is the presence of a wider and at the same time weaker emission component than the classical narrow jet.

## 4. Modeling of Log*N* - Log*S* curve for sGRB

The use of the $\log N$ - $\text{Log}S$ curve to analyze the existence of sGRB emission component other than the classical jet is due to that, unlike the direct approach based on the study of a specific emission mechanism, this method simultaneously covers a large group of explanations, including the existence of a secondary weak emission components. This approach, of course, will not allow us to study in detail the physics of sGRBs, but it will provide an effective technique of measuring the fraction of weak bursts in observations for which the redshift has not been determined. Thus, this approach will make it possible to prove the existence of a population of "weak" short-bursts purely from gamma-ray observations without localizing host galaxies, and therefore without strong selection effects. As part of the modeling, we will consider two approximations, where the second will be some complication of the previous one:

    1) homogeneous jet + homogeneous spherical component;

    2) homogeneous jet + homogeneous cone-shaped component.

## 4. 1. General principles of the model

If to suppose that the GRB source luminosity is described by a normal distribution with some mean, then the theoretical $\log N$ - $\log S$ curve in Euclidean space should have a power-law form with a slope index close to -3/2.

But in reality, the situation is much more complicated. First, let us assume that each merging of two neutron stars results in a sGRB, and the number of sGRBs resulting from a neutron star - black hole merging is negligible relative to the total number of sGRBs, it means that merge rate of binary neutron stars ($MR_{BNS}$) is approximately equal to the rate of sGRB ($R_{sGRB}$), i.e. $MR_{BNS} \approx R_{sGRB}$ (The reverse approach to this problem has already been considered in the work Virgili et al., 2011). Then it is quite correct to assume that the number of such mergers will depend on the average evolutionary state of stars in galaxies. This will be discussed in more detail a little later, but to a first approximation, main contribution to the NS merging give fast mergers, which makes the dependence of the merger frequency on the star formation rate and on the binary system evolution rate significant. Each of these quantities depends on the redshift.

Relativistic effects also make a significant contribution; they are responsible for the appearance of a flat part on the $\log N$ - $\log S$ curve. Let us assume that all spectra of short bursts are described by a power function $\sim E^{-\alpha}$ near given energy range, then, taking cosmological reddening into account, we will already observe a flux:

$$S(z) = \frac{L(1+z)^{-\alpha-1}}{4\pi d_m^2(z)}$$

where $L$ is mean isotropic GRB luminosity, and $d_m$ is accompanying distance.

If we consider a more complex case, it would be necessary to add the function $k(z)$ to the numerator of this expression, which could take into account non-trivial features of the detector sensitivity, but we will take the simple case when the detector has uniform sensitivity in a given energy range, and therefore $k(z) = (1+z)^{-\alpha}$, as written. For definiteness, we will limit ourselves to one cosmological model $\Lambda$CDM, assuming a flat Universe with the parameters: $\Omega_\Lambda = 0.6911 \pm 0.0062$; $\Omega_m = 0.3089 \pm 0.0062$; $H_0 = 67.74 \pm 0.46 \; ^{\text{km}}\!/_{\text{s} \cdot \text{Mps}}$ (Planck Collaboration, 2016).

Then the accompanying distance can be calculated with the use of the formula

$$d_m(z) = \frac{c}{H_0} \int_0^z \frac{dz'}{\sqrt{\Omega_m(1+z')^3 + \Omega_\Lambda}}$$

Having obtained the dependence $S(z)$, for each given $S$ we can estimate a $z$ value beyond which, on average, all sGRBs that have occurred are not visible. Let us know the rate of mergers at redshift $z$ $MR_{BNS}(z)$, then the average number of mergers inside the volume limited by $z(S)$ is given by the integral over the volume, which we reduce to the integral over $z$:

$$N(>S) \sim 4\pi \int_0^{z(S)} d_m^2 \frac{MR_{BNS}(z)}{(1+z)} \frac{d(d_m)}{dz} dz,$$

where the factor $1 + z$ in the denominator appeared due to the discrepancy between the clock rate at the observer and at redshift $z$. $MR_{BNS}(z)$ values can be obtained in different ways, but the simplest is to compare it with the local merger frequency, and then, using this constant, normalize the convolution between the evolution rate (merger probability versus time $f(t)$) and the star formation rate, which should potentially give a value proportional to the rate of mergers

$$MR_{BNS}(z) \sim \int_z^{\infty} SFR(z') \left( f(t(z) - t(z')) \right) \frac{dt}{dz'} dz'$$

If we obtained dependence $N(>S)$ in this way, we find out the average number of mergers that occurred in the volume from which we could see them with a given sensitivity ($S$), but it is also worth considering which of them we can only see as a normal sGRB that part whose jets can hit the Earth, and these are objects of extragalactic nature, then the probability of such an event will be equal to the ratio of the cone angular area to the sphere area $p = (1 - \cos(\theta/2))/2$, where $\theta$ is the jet opening angle. Now we have considered the model for "ordinary" sGRBs, but as already mentioned, we are interested in two-component bursts.

If we are talking about the spherical component, then the probability of observation will become equal to one, but the average luminosity will be significantly lower, that is, the constant in the expression $S(z)$ will change. We will assume that most short bursts will have a weak second component, and therefore there will be no changes in the expressions for the merger rate. A more complex model with the presence of a wider, but still non-spherical component can be considered in a similar way, but the probability in it will be determined in the same way as in the case of a jet, but with a different opening angle $\theta_{wide}$, and minus the probability of getting into the jet itself

$$p = (1 - \cos(\theta_{wide}/2))/2 - (1 - \cos(\theta/2))/2.$$

For each of the models, we will take the same average isotropic luminosity of the faint component, taking as a basis the ratio between the isotropic luminosity $L_{iso}$ of GRB170817A and the average isotropic luminosity $<L_{iso}>$ of short bursts (Fong et al., 2015) ($10^{47}$ and $10^{51}$ erg/s, respectively)

$$L_{iso}(\text{GRB170817A}) \approx <L_{iso}>(\text{sGRB})/10^4.$$

## 4.2. Star formation function

Let us consider the components of our model in a little more detail. The star formation function (SFR) determines how much mass will be converted into stars on average in cubic Mpc per year. Obviously, part of this mass will be spent on the creation of stars, which, as a result of evolution, will turn into a close system of two NSs and give us a merger that can be detected as a sGRB. Hence, it seems plausible that the rate of mergers is proportional to the rate of star formation, taking into account a certain time delay.

As a unified law of SFR change on $z$ (SFR($z$)), we will take the so-called Madau function (Madau & Dickinson, 2014), which empirically describes average star formation in the Universe up to $z = 8$. It is based on multi-wave observations in the FUV (Far Ultra Violet), MIR (Mid Infra Red) and FIR (Far Infra Red) ranges. Other studies of nebula in lines or radio emission are also important, but they provide more limited or indirect information. An example of obtained smooth Madau function can be seen in Fig. 3.

It is worth considering that, using the Madau function, we agree with another strong assumption, i.e. the initial mass function weakly depends on the redshift, which, upon closer examination, is obviously not true, since the chemical composition changes during the evolution process, and therefore the forming stars change. Due to the fact that we do not have enough information to describe the change in the initial mass function at large $z$, we will take the unmodified Madau function as the best approximation of the actually observed picture. The empirical form of the function is given by the equation

$$SFR(z) = 0.015 \frac{(1+z)^{2.7}}{1+[(1+z)/2.9]^{5.6}} \, \text{M}_\odot \text{ year}^{-1} \text{Mpc}^{-3},$$

just this function is presented in Fig. 3.

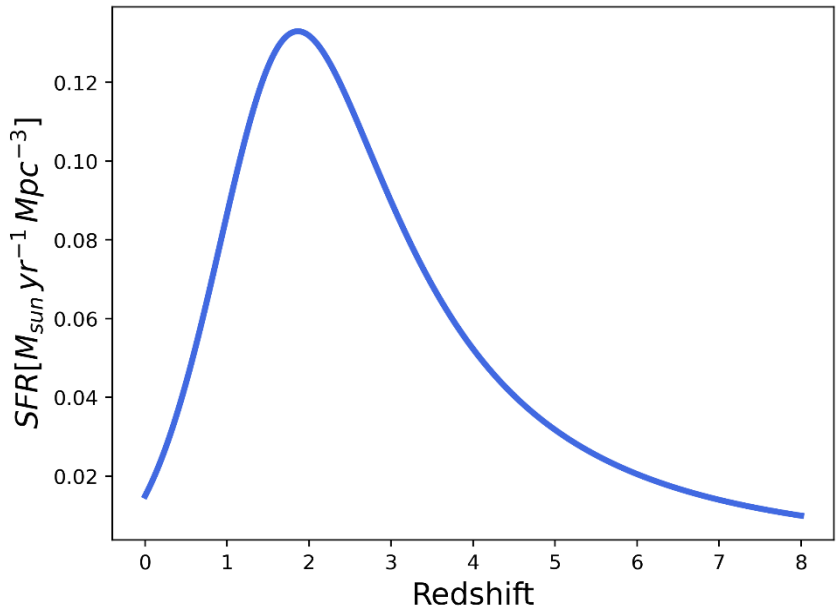

Fig. 3: Type of star formation function used in the model.

### 4.3 Neutron star merging rate

As we have already discussed, the star formation rate by itself is not enough. It is important to find on what it should be normalized and take into account the time delay between the formation of a NS pair and its merging. The Scenario Machine can answer these two questions at once (Lipunov et al., 1995a, Lipunov et al., 1996) The Scenario Machine method to calculate evolution of binary stars is basically Monte-Carlo method for statistical simulation of large ensembles of stellar binary systems originally posted by Kornilov & Lipunov (1983) for binaries. The method is based on the construction of a great number of single evolutionary tracks with different initial conditions.

Taking the normalized merger probability function from the scenario engine as $f(t)$, the merger rate can be calculated numerically as a function of $z$:

$$MR_{BNS}(z) \sim \int_z^\infty SFR(z') \left( f(t(z) - t(z')) \frac{dt}{dz'} dz' \right.$$

To integrate this expression, it is necessary to use the relationship between $t$ and $z$, which, in the framework of a flat $\Lambda$CDM model, is given by a set of equations:

$$H(z) = H_0 \sqrt{\Omega_m (1+z)^3 + \Omega_\Lambda}$$

$$T(z) = \frac{1}{H_0} \int_0^z \frac{H_0}{(1+z')H(z')} dz'$$

This equation can be solved analytically

$$z(T) = \left(\frac{\Omega_\Lambda}{\Omega_m}\right)^{\frac{1}{3}} \left[\left(\frac{1 + W(T)}{1 - W(T)}\right)^2 - 1\right]^{\frac{1}{3}} - 1,$$

where

$$W(T) = \exp\left[\ln\left(\frac{1 + \sqrt{\Omega_\Lambda}}{1 - \sqrt{\Omega_\Lambda}}\right) - 3H_0\sqrt{\Omega_\Lambda}T\right]$$

Using a Scenario Machine is not the only option here. As other studies show (Wanderman, Piran, 2015), there are two functional approximations for the merging delay time:

1) power law $f(\tau) = \tau^{-\alpha_t}$,, where $\tau$ starts at 20 million years, i.e. the minimum period of system evolution, which is also reflected in the Scenario Machine curve, the simplest version of this relationship $\alpha_t = 1$ (Piran 1992a) gives a divergent solution, the Scenario Machine approximation gives values of $\alpha_t \approx 1.5$, which gives the delay for more than 60% of the bursts to less than 100 million years. If $\alpha_t > 3$, then $MR_{BNS}(z) \sim \mathrm{SFR}(z)$;

2) lognormal distribution with a typical width of the time delay $\sigma_t$: $f(\tau) = \exp\left(-\frac{(\ln\tau - \ln t_d)^2}{2\sigma_t^2}\right) /$ $(\sqrt{2\pi}\sigma_t)$, such a model with a small width, in fact, reflects a constant delay.

The last point which should be discussed before modeling is the normalization of the burst number. The simplest variant is to tie the local burst frequency in a linear part of log$N$ – log$S$ curve to the observed one, thereby we will not only obtain the most plausible parameters, but will also be able to take into account part of the systematic errors. To do this, the log$N$ -log$S$ curve should be divided on the number of years of observations, which for the Fermi GBM is about 10 years. It is worth noting that this is only a lower estimation of the number of real events due to the fact that for a significant time instrument such as Fermi GBM, for technical reasons, are not able to survey the sky with sufficient sensitivity. But there is also a second option for solving this problem, i.e. to use the local merger frequency, which we extrapolate using SFR and $f(\tau)$. There are several estimates of local merger rates, but the most plausible values seem to be around ~$10^3$ mergers per Gigaparsec per year (Mandel, Broekgaarden, 2022), which is also consistent with estimates derived from GW observations (Abbott et al., 2023, Lipunov et al., 2022).

## 5. Results of modeling. Parameter limits.

This section will present the main modeling results and conclusions to them. The variable parameters were the jet opening angle, the opening angle of the faint component, and the base frequency of NS mergers.

Let us consider log$N$ - log$S$ curves for different opening angles of the main jet and the spherical component with isotropic luminosity of the GRB170817A level (see Fig. 4).

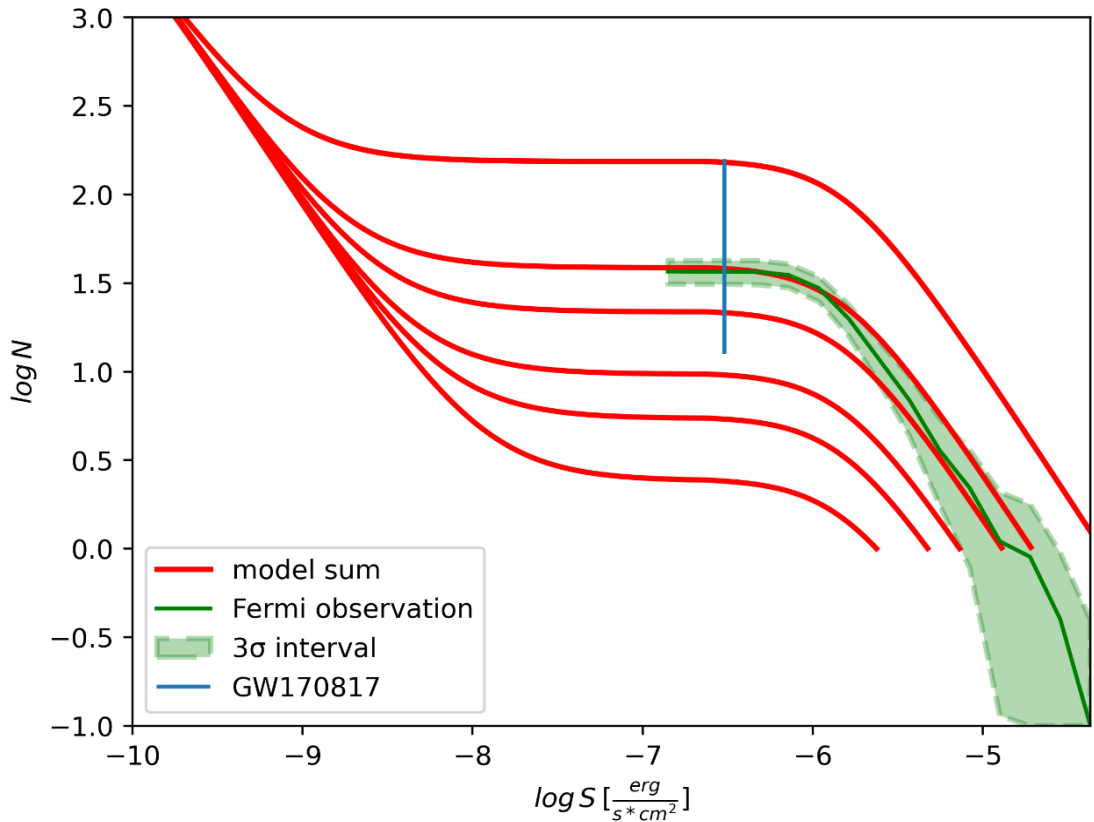

Fig.4. Curves log$N$ - log$S$ with fixed NS + NS merging frequency for different jet opening angles, down up 4°, 6°, 8°, 12°, 16°, 32°. Blue line indicates GW170817 position.

In these curves, the burst frequency is fixed from the local rate of NS mergers, which is on the order of about 10 mergers per year per cubic Gpc, which is significantly lower than the estimates we showed in the Section 3.3. It is clear that it is possible to get into the observed range by reducing the opening angle of the classical jet and at the same time increasing proportionally the merging rate, then assuming that the detection efficiency is equal to 1, we can say that $\frac{1-\cos\left(\frac{\theta}{2}\right)}{2} * MR_{BNS}(\frac{1}{\text{год } Gpc^3}) = 0.05$. Hence, assuming that the merger rate is $10^3$, we obtain that the jet average opening angle must be greater than 2°, otherwise there will not be enough mergers to explain the majority of short bursts. This constraint is consistent with the opening angle measured from sGRB afterglows (Troja et al. 2016). Note that the product described above was obtained under the assumption of 100% efficiency of the gamma-ray detector, which, obviously, is not actually true, and therefore, our model can potentially satisfy a larger product of the solid angle and the merger rate. Curves in Fig. 4 also show that even at very small opening angles, the rise from the second component will become noticeable only when the sensitivity approaches $10^{-8}$ erg/cm$^2$.

Let us fix the product of the solid angle and the merger local rate and see how the picture would change if the second component were not spherical (see Fig. 5).

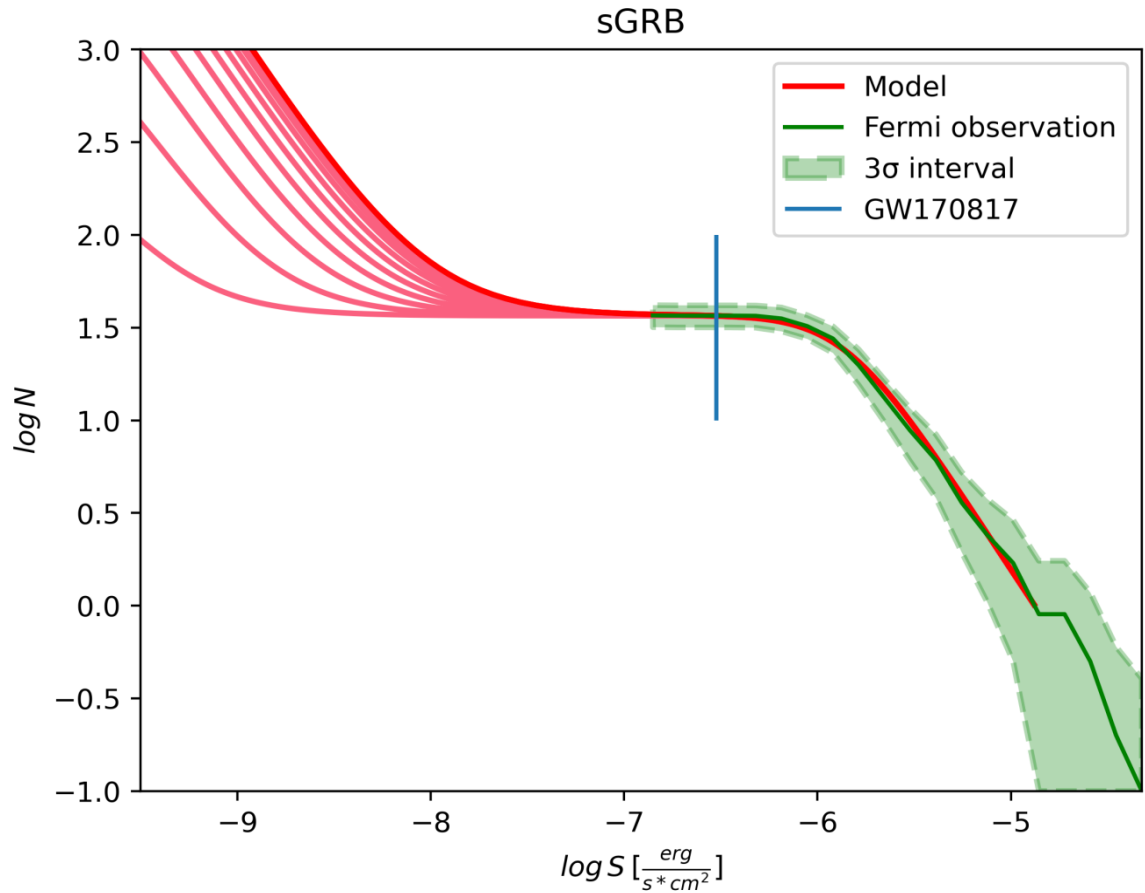

Fig. 5. Curves $\log N$ - $\log S$ for main jet opening angle 8°, $MR_{BNS} \sim 10^2$, different opening angles of second component, down up from 24° to 360° (quasi-spherical case).

The smaller the coverage area of the second component, the lower $S$ values by which the second component will make significant contribution to the $\log N$ - $\log S$ curve. Immediately we can see another effect, i.e. decrease of the second component coverage area makes the observation of GW170817 itself very unlikely. So, for a weak component with opening angles less than 60° and for a jet opening angle approximately equals 8°, the probability of observing the GW170817 event will be less than 1 time in 300 years, which seems implausible. Let us consider this problem in more detail, for this we will highlight the contribution of the secondary component to the $\log N$ - $\log S$ curve (see Fig. 6).

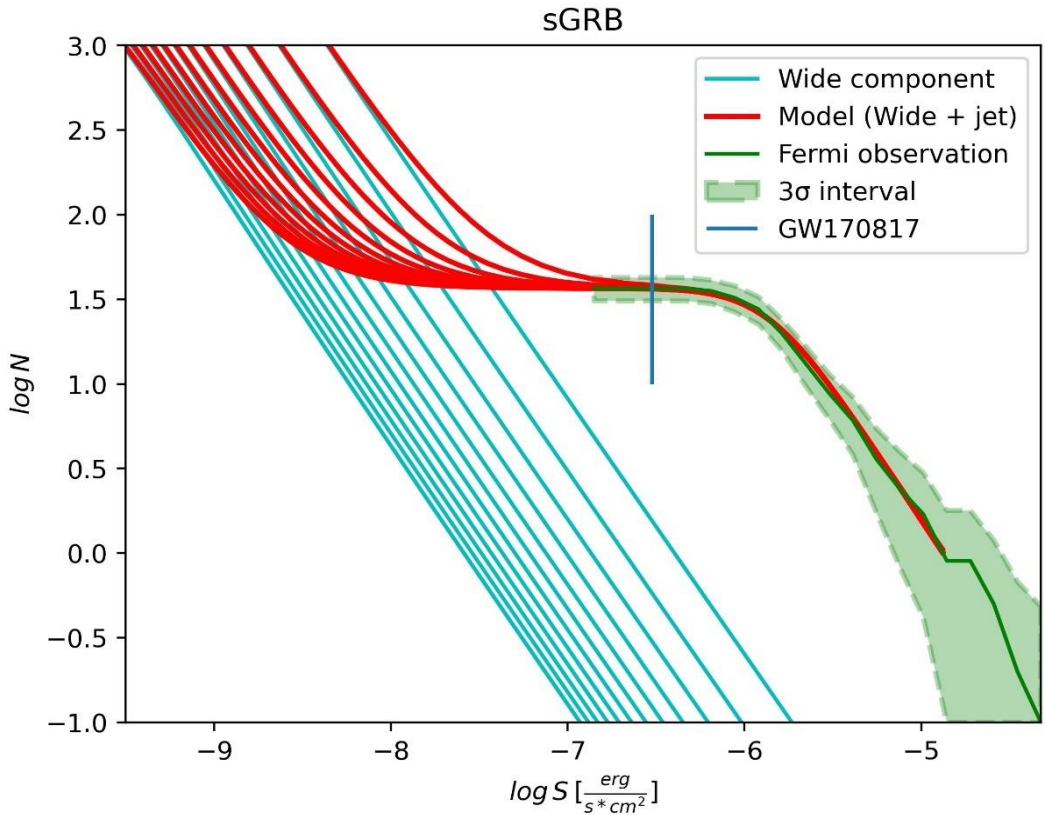

Fig. 6: Curves log*N* - log*S* with a fixed product of the NS + NS merger frequency and the coverage area for different jet opening angles, from left to right from 2° to 16°, the merger rate is from 600 to 10 mergers per year per cubic Gpc. Thin blue lines show the log*N* - log*S* curves for the weak component.

The intersection of the lines corresponding to the second component with the horizontal straight line log*N* = 1 allows us to obtain the typical value of the flux at which the second component becomes noticeable. The intersection with the vertical line corresponding to GW170817 allows us to estimate the probability of the GW170817 event, as well as estimate the likely number of objects similar to this burst. For our simple models, we can observe that only within models with a small average jet opening angle it would actually be possible to observe the GW170817 event with a probability higher than once every 10 years, so we can accurately say that the two-component emission model at adequate values the opening angle of the main jet predicts a low probability of finding an event similar to GW170817 in already observed sGRBs, while intersections with the log*N* = 1 line show that with increasing sensitivity to $10^{-8}$ erg/s cm², the number of observed second components will become significant and we can clearly decide which scenario is implemented in GW170817. It is important to note that an upper limit on the number of nearby sGRB can be obtained by cross-correlation between bursts position and nearby galaxies(Dichiara et al., 2020).

# 6. Methods of GRB detector sensitivity improvement

Traditionally, to detect GRBs relatively cheap scintillation detectors are used, which can be manufactured with a sufficiently large area (on the order of hundreds of cm²) and thickness (several cm). As a rule, events are selected in the energy range of detected quanta from 50 to 300 keV. A typical background count in detector channels corresponding to the given range is hundreds of Hz, i.e. the background count rate per unit area ($N_f$) is ~1 - 2 counts/cm$^2$s. Then, assuming the Poisson nature of the background count distribution, we obtain for a detector with an effective area $S$ the value of the standard deviation of the background count ($\sigma$):

$$\sigma = \sqrt{N_f S \Delta t}, \tag{5.1}$$

where $\Delta t$ is the selected burst duration.

If we take the average energy of the detected gamma quantum as 100 keV, the burst duration $\Delta t$ = 1 s, the background count rate as 2 counts/cm$^2$s, then for a detector with an area of 500 cm$^2$ at a confidence level of $5\sigma$ we obtain an estimate of the minimum detected total flux (fluence) F in the indicated energy range ~ $5\cdot10^{-8}$ erg/cm$^2$. Since for most bursts the energy spectra extend beyond the above range, the average gamma ray energy will be more than 100 keV, respectively, the real sensitivity threshold for long-lasting gamma bursts is ~$10^{-7}$ erg/cm$^2$. This value is typical for most orbital experiments for observing GRB, such as BATSE CGRO, GBM Fermi, etc.

The detector sensitivity, as follows from (5.1), can be improved by increasing the effective area and reducing the background. For sGRBs, which are associated with the merger of relativistic compact objects, the typical energy of emitted photons $E_0$ lies in the range from 100 to 600 keV with a weighted average value of ~250 keV. Taking the approximation of the spectrum in the form of a power law with exponential cut-off:

$$J(E) = J_0 E^{-\gamma} \exp(-E/E_0), \tag{5.2}$$

where $J_0$ is the normalization coefficient, we find that to measure the total fluence per burst it is practically sufficient to detect gamma quanta with energies up to $E_0$.

The most popular scintillators used to detect GRBs are NaI(Tl), CsI(Tl). To ensure acceptable detection efficiency (at the level of 60%), their thickness must be at least 2 cm. As experience of experiments on the Vernov (Panasyuk et al., 2016) and Lomonosov (Sadovnichii et al., 2017) satellites shows, the minimum background in crystals of this thickness (i.e. in equatorial regions) is exactly ~2 count/cm$^2$s. It can be reduced by using active shield. Thus, in particular, on the Lomonosov satellite, so-called phoswitch detectors based on the NaI(Tl)/CsI(Tl) combination were used to detect GRBs. In that case, NaI(Tl) crystals od 0.3 cm thick were used as the main detecting element, and CsI(Tl) crystals of 1.7 cm thick were used as active shield. This method made it possible to reduce the background in NaI(Tl) detectors to values less than 1 count/cm$^2$s in equatorial regions. However, a NaI(Tl) thickness of 0.3 cm provides sufficient efficiency only up to energies of ~200 keV.

In the case of spectral representation (5.2), we can assume that up to energies $E_0$ it is possible to use a power-law approximation of the energy spectrum. In this case, the relation for the average energy is known:

$$\overline{E} = \left[ \frac{\gamma(E_2 - E_1)}{E_1^{-\gamma} - E_2^{-\gamma}} \right]^{1/\gamma + 1}. \tag{5.3}$$

In (5.3) $E_1$, $E_2$ are the energy range boundaries. Taking $E_1 = 20$ keV, $E_2 = 500$ keV, we obtain the average photon energy values 73 keV for the power index $\gamma = 2$, 50 keV for $\gamma = 4$. From this it can be seen that the average photon energy does not greatly depend on the energy spectrum index. Let us take this energy equal to 70 keV, then for a detector with an area of 500 cm$^2$ and a background count of 2 count/cm$^2$s, the threshold value of the fluence of a burst with a duration of 1 s, detected at the $5\sigma$ significance level will be ~$3.5\cdot10^{-8}$ erg/cm$^2$.

To improve the sensitivity to ~$10^{-8}$ erg/cm$^2$, it is necessary either to increase the effective area of the detector by 12.25 times, or to reduce the background count level by a corresponding number of times. It seems optimal to increase the effective detector area to 1600 - 2000 cm$^2$ and, accordingly, reduce the background by 3-4 times - to 0.5 - 0.7 counts/cm$^2$s. The detector of this area remains quite compact and can be placed on mini-satellites. In terms of background reduction, this can be achieved by using active shielding in combination with advanced scintillation detectors such as Ce:GAGG or YSO. It was noted above that the Lomonosov satellite used a NaI(Tl)/CsI(Tl) phoswitch detector to detect GRBs, in which a background level of ~0.75 counts/cm$^2$s was achieved in a NaI(Tl) crystal of 0.3 cm thick, but at the same time a sufficiently high efficiency of detecting of gamma rays with energies above 200 keV was not ensured. If a Ce:GAGG scintillator with a thickness of 0.5 cm is used as the main detecting element, then due to the fact that it is characterized by a higher density (6.6 g/cm$^3$) and consists of elements with large Z values, acceptable efficiency will be ensured for detecting gamma quanta with energies 500 keV at a level of ~40%. Moreover, since the background in a crystal is determined to a large extent by its volume, when using active shield we can expect its value to be no worse than that achieved on the Lomonosov satellite.

It seems promising to use a pixelated detector, which has a number of important advantages, in particular, it will significantly expand the dynamic range of the instrument, identify detected particles, including separating events associated with the detection of GRBs and magnetospheric electron precipitation, as well as carry out polarization measurements. The modular structure of the instrument's construction makes it possible to mount it to the spacecraft (identical modules can be used to assemble a detector of any given area).

Separately, the possibility of GRB observing on CubeSat spacecraft should be considered. Only very compact instruments can be placed on such satellites; even on a 16U format sutellite it is difficult to accommodate a detector with an effective area of more than 800 cm$^2$. However, cubesat satellites have a number of advantages, in particular, due to their low mass, the background count in instruments placed on such spacecraft will be determined only by the detector's own background. Therefore, for background rejection on cubesats, it makes no sense to use active protection in the detectors. The experience of using space radiation detectors on cubesat satellites of the multi-satellite constellation of Moscow University has shown that detectors based on a 1 cm thick CsI(Tl) scintillator provide a background count of ~1 count/cm$^2$s (Svertilov et al., 2020). If a Ce:GAGG scintillator with a thickness of $0.1 - 0.2$ cm is used as a detecting element, the expected background count level in it can be ~0.5 counts/cm$^2$s. In the case of using a YSO scintillator, which is weakly activated by external radiation, an even lower background count can be expected. Thus, if a detector with an effective area of ~400 cm$^2$ based on the YSO scintillator is used on a cubesat format satellite, the gamma-ray burst detection threshold at a level of ~2·10$^{-8}$ erg/cm$^2$ can be achieved. It should also be noted that the "√2" factor can still be gained in the fluence of sGRBs, since their weighted average duration is approximately 0.5 s, while the above estimates were made for bursts with a duration of 1 s.

With the above-mentioned scintillator thicknesses, it is difficult to ensure a sufficiently high detection efficiency of gamma rays with energies >100 keV. However, the energy range of detected quanta without loss of sensitivity can be expanded if a combined, so-called "foswitch" detector is used, for example, based on a combination of a thin $(0.1 - 0.2$ cm thick) Ce:GAGG or YSO scintillator and a CsI(Tl) crystal of 2 cm thick or more. The use of a CsI(Tl) scintillator will not only expand the energy range of detected gamma rays towards higher energies, but will also further reduce the background in Ce:GAGG or YSO scintillators, since the CsI(Tl) crystal can be used as active shield. In this case, to pick up signals from both scintillators, it can be uses the same photodetectors and signals from them can be separated with the use appropriate electronics, i.e. signal shape form analyzer, because photodetector output signals will differ significantly for "fast" scintillators such as Ce:GAGG or YSO with a short illumination and a "slower" scintillator CsI(Tl).

## 7. Discussion

The signal/noise ratio and, respectively, the sensitivity to detecting weak GRBs can be significantly increased if the selection of events is not carried out by the full fluence (or by the fluence in a wide energy range), as has been done so far in most experiments on observing GRBs, but by the spectral flux density $J$ (measured in photo/cm²keV) or by the total flux in a narrow energy range. The optimal range in this regard seems to be 25 – 50 keV. For example, in the GRIF experiment with the RX-2 X-ray instrument at the Mir orbital station (Kudryavtsev et al., 1998), in this energy range a background count level of ~0.5 counts/cm²s was achieved, which made it possible to detect bursts with total fluxes of ~3·10⁻⁸ erg/cm².

From this point the expected $\log N$ – $\log S$ distribution was recalculated for fluxes in the energy range 25 – 50 keV. The corresponding distribution is shown in Fig. 7. The resulting curve shows the ideal case if we observe in a narrow band (25 - 50 keV) with an ideal spectrum of the exponentially cut off power law type, with average parameter values for sGRBs. Since the actual spectrum of the broad component may not be well described by the model, there may be a relatively small systematic bias in $\log S$ for the broad component.

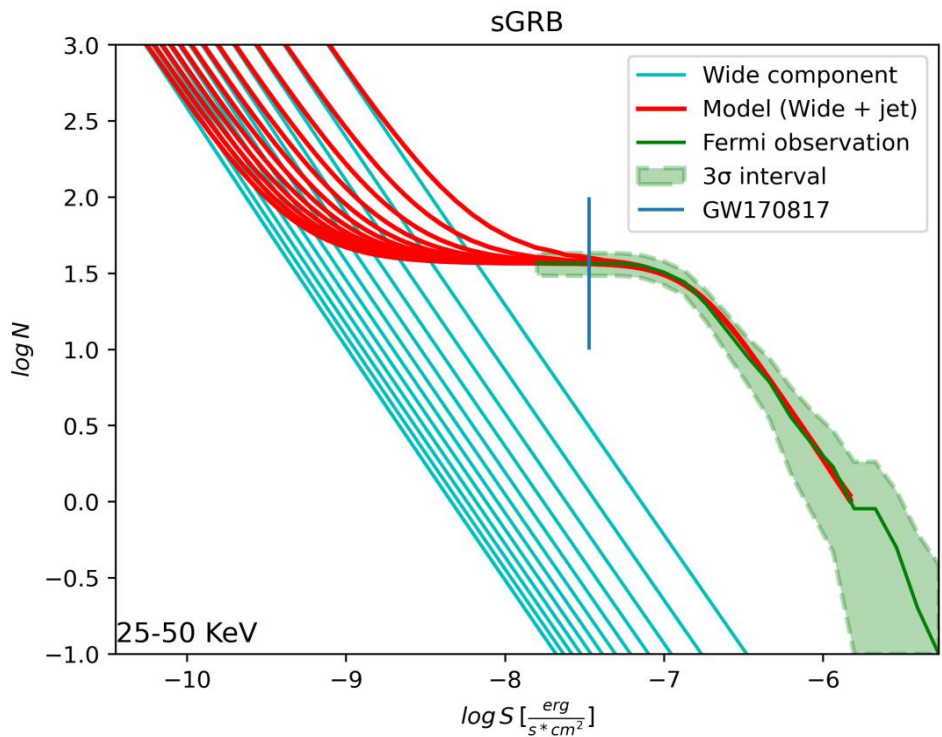

Fig. 7. Curves $\log N$ - $\log S$ as the same in Fig. 6 recalculated for $S$ values in 25 – 50 keV range. The red lines correspond to different opening angles of the main jet component from 16 to 2 degrees, from left to right.

As follows from Fig. 7, a "plato" in the $\log N$ – $\log S$ distribution is observed at $S \sim 10^{-7}$ - $5\cdot10^{-8}$ erg/cm². If we provide a background count rate of the detector in a 25-50 keV channel at a level of ~0.2 - 0.5 counts/cm²s, which, as noted above, is quite achievable even on Cubesat satellites using advanced scintillators of the Ce:GAGG or YSO type, then for detectors with an effective area ~400 cm², sensitivity can be ensured at a level of $5\cdot10^{-9}$ erg/cm2 in the mentioned above energy range. As can be seen from Fig. 7, this makes it possible to "feel" the rise in the $N(>S)$ distribution due to the detection of a weak homogeneous component of the emission of sGRBs.

# 8. Conclusions

In conclusion, the following should be noted. GRB 170817A (GW 170817) is a unique object among sGRBs, first of all because we know the distance to the source quite accurately, and secondly, because this distance is extremely small compared to any other GRBs. Assuming that the low energy of this event can be explained by the fact that we are not in the jet cone, but observing either a quasi-spherical component of the sGRB emission, or some wider angular jet structure with a sharp transition in luminosity inside. We cannot currently conclude which scenario corresponds to the GW170817 event, but modeling of the $\log N$ - $\log S$ curve allows us to construct a criterion for future increases in the sensitivity of gamma-ray detectors, which will unambiguously resolve the question of the existence of a second emission component in sGRBst. Having carried out statistical calculations and using the known local density of matter and estimates of the merger frequencies NSs, which are progenitors of sGRBs, an estimate was obtained for the sensitivity necessary to observe an increase in the number of bursts with a decrease in $S$.

In the most optimistic scenarios, a completely homogeneous spherical component with a small average jet angle opening ($\sim 4°$) with a typical luminosity like that of 170817, will become a noticeable fraction only when a stable limit of observed fluences is obtained at $5 \cdot 10^{-8}$ erg/s cm$^2$ (in a wide energy range of detected gamma rays) and $5 \cdot 10^{-9}$ erg/cm$^2$ (in the range of detected gamma quanta 25-50 keV), which is theoretically achievable for new advanced gamma detectors. With adequate restrictions on the model, it is possible to exclude a large fraction of sGRBs similar to GW170817 in current observations for which redshifts were not obtained, that is, within the framework of the faint secondary component model, the GW170817 event itself is quite rare (no more than 1 event per year).

**Acknowledgments**

Topolev V.V. acknowledges the support from the Theoretical Physics and Mathematics Advancement Foundation "BASIS" (23-2-10-35-1). MASTER data for GRB170817 was received with Lomonosov MSU development program in 2017 year.